\documentclass[vecphys]{svmult}

\usepackage{makeidx}         
\usepackage{graphicx}        
\usepackage{multicol}        
\usepackage[bottom]{footmisc}

\makeindex             

\usepackage{graphicx}

\begin{document}

\title*{Six Hot Topics in Planetary Astronomy} 

\author{David Jewitt}

\institute{Institute for Astronomy, University of Hawaii, 2680 Woodlawn Drive, Honolulu, HI 96822, USA
\texttt{jewitt@hawaii.edu} 
}
%
%
\maketitle

\begin{center}\bf\large
Abstract
\end{center}

Six hot topics in modern planetary astronomy are described: 1) lightcurves and densities of small bodies 2) colors of Kuiper belt objects and the distribution of the ultrared matter 3) spectroscopy and the crystallinity of ice in the outer Solar system 4) irregular satellites of the giant planets 5) the Main Belt Comets and 6) comets and meteor stream parents.

\section{Introduction}

The direction given to the authors of this book\footnote{Article completed in January 2008 and to be published in ``Small Bodies in Planetary Systems'' (Mann, Nakamura and Mukai, editors), Lecture Notes in Physics, Volume 758, Springer Academic Publishers, 2008.  Funded by the Japanese 21st Century Centers of Excellence Program``Origin and Evolution of Planetary System''.}
 is to show some of the exciting recent developments in the study of the Solar system.  Of course, ``exciting'' is a subjective term, and one which gives this author a lot of latitude.  The most exciting science 
subjects for me are the ones I am working on, so I 
have written this chapter as a series of vignettes describing six topics from my own on-going research and from the research of
my students and colleagues [principally Henry Hsieh (main belt comets), Bin Yang (spectra), Jane Luu (colors and spectra), Scott Sheppard (irregular satellites and lightcurves), Pedro Lacerda (lightcurves), Nuno Peixinho (colors) and Toshi Kasuga (meteors)].
What follows is not so much a review as a window onto these six, particularly active parts of
modern planetary astronomy.  The reader who wants the raw science or access to the full literature on a given subject has only to go to the
journals or to astro-ph: the internet makes it easy.  My objective here is to focus attention mainly on newer, perhaps less-known work, the big-picture significance of which has yet to become clear.  Relevant questions are listed explicitly where they crop up in each section of the text.

Research in modern planetary astronomy is concentrated on the small bodies of the Solar system rather than on, as in the past, the major planets.  This is because the small bodies are relatively unstudied and much of what we find out about them is new and surprising.  In fact, many of the different populations of small bodies have only recently been discovered (the Kuiper belt and the Main Belt Comets are good examples) and few have much prospect of being investigated, close-up, by spacecraft in the foreseeable future.  Observations with telescopes are the main practical way to learn about these objects.

\begin{figure}
\centering
\includegraphics[height=9cm]{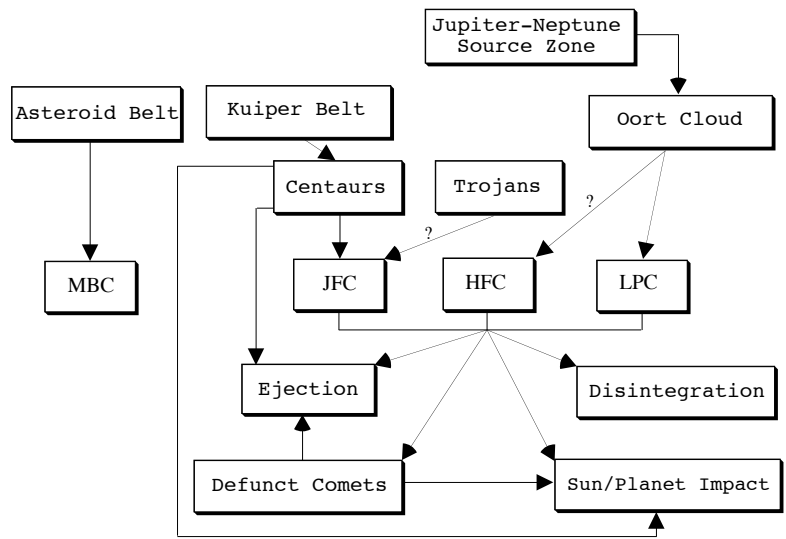}
%
%
\caption{Schematic showing the connections between some of the Solar system's small body populations.  Acronyms are MBC: Main-belt comet, JFC: Jupiter family comet, HFC: Halley family comet, LPC: long-period comet.   Arrows mark interrelations.  For example, the Kuiper belt feeds the Centaurs which become relabeled as JFCs when dynamically interacting with Jupiter.  Most Centaurs die by being ejected from the Solar system or by striking a planet or the Sun.  The JFCs die by one or more of four labeled processes.   Arrows marked ``?'' show connections that remain uncertain.  Loss processes for the MBCs are not yet known.  Figure from \cite{jew07}.}

\label{layout}       
\end{figure}

The first necessary step in this chapter is to lay out the small bodies of the Solar system in a clear way, so that we know what we are talking about.  This is done in Figure \ref{layout}.  There, the main source regions (asteroid belt, Kuiper belt\index{Kuiper belt} and Oort cloud\index{Oort cloud}) are shown at the top of the diagram.  Objects now in the 50,000 AU scale Oort cloud were formed in the Jupiter-Neptune zone and then scattered outwards by strong planetary perturbations.  Their perihelia were lifted by torques from passing stars and from the galactic tide.    Bodies deflected back into the planetary region from the Oort cloud are labelled long period comets (LPCs)\index{Long period comets}, distinguished by large, weakly bound and isotropically distributed orbits.  Halley family comets (HFCs)\index{Halley family comets} have smaller orbits that are more often prograde than retrograde.  Their source has not been established but is likely to lie in the inner regions of Oort's cloud. [The long-period and Halley family comets are sometimes lumped together and given the mangled-English label ``nearly isotropic comets'', by which it is meant that the lines of apsides of the orbits of these bodies are nearly isotropically distributed.] Jupiter family comets (JFCs) have small semimajor axes, inclinations and eccentricities and dynamics controlled by strong interactions with Jupiter.  Their source is thought to be somewhere in the Kuiper belt, but it is not clear which regions of the Kuiper belt actually supply the comets.   Before they are trapped by Jupiter and while they are strongly scattered by the giant planets, escaped Kuiper belt objects are labeled Centaurs.\index{Centaurs}  [The Centaurs and JFCs typically possess modest orbital inclinations and are sometimes referred to as members of the ``ecliptic comet'' group for this reason.]  The most recently discovered comets are the ice-rich asteroids (or main-belt comets, MBCs) \index{Main-belt comets}probably formed in-place at $\sim$ 3 AU.  They do not seem to interact with the other populations and therefore constitute the third-known cometary reservoir, after the Oort cloud and the Kuiper belt.  Trojan\index{Trojan asteroids} ``asteroids'' are likely ice-rich bodies stabilized in the 1:1 mean motion resonances of the planets (Trojans of both Jupiter and Neptune \cite{she06} have been found).  The locations of their origin are unknown.  Comets ``die'' most commonly by being ejected from the Solar system.  Those not ejected disintegrate, devolatilize or impact the planets or the Sun. 

Research is on-going into every box in Figure \ref{layout} and into the arrows that symbolize the relationships between the objects in the boxes.  Indeed, the key advance of the past one and a half decades is that we now clearly see both the boxes \textit{and} the relationships that exist between them.  In this sense, the six hot topics of this chapter are really one: we aim to trace the different kinds of small body populations back to their sources and so to better understand the origin of the entire Solar system.

\begin{figure}
\centering
\includegraphics[height=9cm]{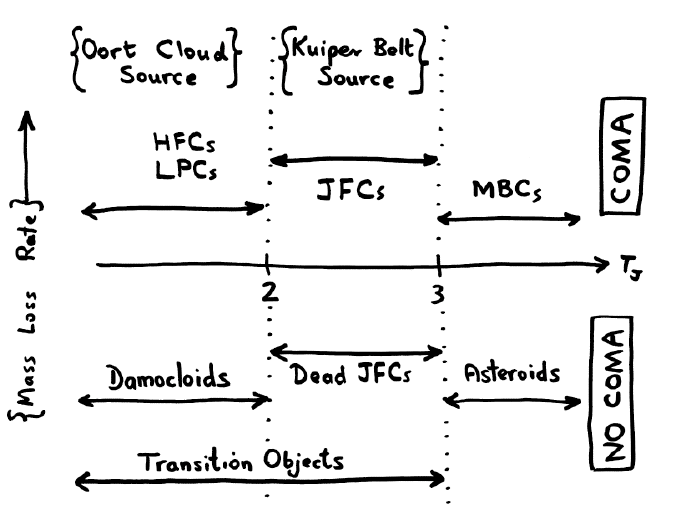}
%
%
\caption{Two parameter classification of some of the small-body populations discussed here.  In the horizontal direction, objects are classified by their Tisserand parameter \index{Tisserand parameter}measured with respect to Jupiter (dynamical comets have $T_J <$ 3, dynamical asteroids $T_J > $ 3). In the vertical direction, objects are classified by whether or not they show evidence for mass loss, presumed to be driven by the sublimation of near-surface volatiles.  Objects above the line are observationally comets because they show comae and/or tails. while objects below the line are observationally inactive and so classified as asteroids.  Ideally, objects should be placed vertically in this diagram based upon measurements of their mass loss rates.  Given our limited knowledge, however, it is more practical at present to use a ``one bit'' classification in which objects are either measurably active or not.}

\label{schema}       
\end{figure}

A second schematic (Figure \ref{schema}) attempts to clarify some of the small-body nomenclature.  It shows a two-parameter classification, reflecting the fact that both dynamical properties and physical properties are regularly used to label objects in the Solar system.  The horizontal axis in Figure \ref{schema} is the Tisserand parameter measured with respect to Jupiter.  This is defined by 

\begin{equation}
T_J = \frac{a_J}{a} + 2\left((1-e^2)\frac{a}{a_J}\right)^{1/2}\cos(i)
\end{equation}

\noindent where $a$, $e$ and $i$ are the semimajor axis, eccentricity and
inclination of the orbit while $a_J$ = 5.2 AU is the semimajor axis of
the orbit of Jupiter.  The Tisserand parameter provides a measure of the relative
velocity of approach to Jupiter: Jupiter itself has $T_J$ = 3, most
comets have $T_J <$ 3 while main-belt asteroids generally have $T_J >$ 3.

The position of an object either above or below the x-axis in Figure \ref{schema} shows whether the object has a measurable coma (gravitationally unbound atmosphere) or not.  The presence of a coma is related, in an unclear way, to the presence of near-surface volatiles.  
Objects showing comae are, by the physical definition of the word, ``comets''.    The JFCs \index{Jupiter family comets}are those comets with  2 $< T_J \le$ 3.   Non-outgassing objects with 2 $< T_J \le$ 3 are called Transition Objects (TOs), \index{Transition objects}or sometimes ``dead comets'' or ``dormant comets''.  Comets with $T_J \le$ 2 fall into the LPC and HFC comet types.   Non-outgassing objects with $T_J \le$ 2 are called ``Damocloids'': \index{Damocloids}their orbital elements suggest that most are the dead or dormant nuclei of HFCs \cite{jew05}.   The MBCs are like asteroids in having
 $T_J >$ 3 but differ in showing comae.

\section{Lightcurves and Densities}
\label{densities}
Lightcurves \index{Lightcurves}offer valuable opportunities to assess the shapes and rotational states of bodies that are, generally, too small 
in angular extent to be resolved, even with the best existing adaptive optics systems (which currently offer resolution $\sim$0.05 arcsec).  
It is also possible, at least for some objects, to use lightcurves to estimate the bulk density of a body.

\subsection{Lightcurves}
The first thing to acknowledge is that there are no \textit{unique} interpretations of lightcurves.  Rotational variability in the scattered light is influenced by the shape of the body, by the surface distribution of materials having different albedos, by the surface scattering function, the viewing geometry and so on.  This non-uniqueness is unarguable, as it was when first noted as long ago as 1906 \cite{rus06}.  One hundred years later, the uniqueness problem is still dredged up by critics in response to new work.  But, while no mathematically rigorous proof exists that a given lightcurve can be interpreted in any particular way, there is a large and growing body of exciting and illuminating work based on rotational lightcurves of small bodies.  This is possible because, wherever supplementary information is available, we find that the lightcurves of small Solar system bodies, almost without exception, are dominated by rotational modulation of the projected cross-section rather than by spatial variations in the albedo.  That is to say, most of the available evidence shows that albedo non-uniformity is small (the exceptions tend to be pathological, like Saturn's two-faced, synchronous satellite Iapetus, and not of general relevance to objects in heliocentric orbits).  Rotational modulation of the projected cross-section (body shape) determines most lightcurves.

What controls the body shape?  Sufficiently strong bodies can maintain any shape against their own gravity, but evidence from the study of main-belt and near-Earth asteroids shows that large bodies are not strong.  Their interiors have been fractured and weakened by past impacts (in the case of the weakly agglomerated comets, the interior strengths may have been small to start with).   In the limiting case of zero strength, the shape of a body must relax to an equilibrium configuration that is a function of the body density and angular momentum.  These equilibrium shapes  follow a well-defined progression from spheres (no rotation) to oblate spheroids (bodies flattened along the polar direction, known as Maclaurin spheroids) to tri-axial figures (the Jacobi ellipsoids that grow longer up to a critical angular momentum content above which no single-body equilibrium shape exists).  Single, strengthless bodies with specific angular momenta higher than a critical value (that depends only on the density) are unstable to rotational fission. Chandrasekhar \cite{cha87} famously calculated these shapes.  

Under the \textit{assumption} of zero strength, the shape and rotation of a body can thus be used to estimate the density.  The validity of the assumption is, of course, questionable and good reasons to doubt the zero-strength assumption exist.  After all, small Solar system bodies are rocks, not liquids, and so they cannot literally be strengthless, especially in compression.  Even if they lack overall tensile or cohesive strength, pressure-induced shear strength between components gravitationally bound in an aggregate should inhibit complete relaxation to the equilibrium state, much as grains of sand in a pile do not flow under gravity like a liquid because of frictional forces between the grains \cite{hol07}.

\begin{figure}
\centering
\includegraphics[height=8cm]{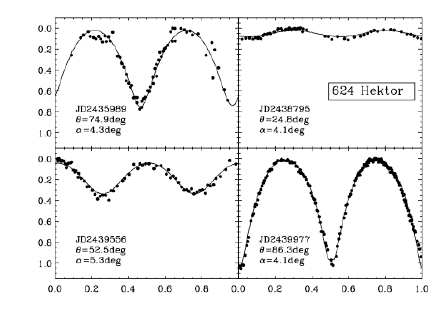}
%
%
\caption{Lightcurves of (624) Hektor at four aspect angles (the angle between the line of sight and the spin direction) compared with an equilibrium binary model. In each panel, the x-axis displays rotational phase (computed for period = 6.9 hr) and the y-axis shows the relative magnitude.  The fits provide a remarkably good representation of the data, lending credibility to the model.  Figure from \cite{lj07}.}
\label{hektor}       
\end{figure}

Despite these legitimate reservations, the evidence suggests that equilibrium models can indeed work very well when the bodies and their lightcurve ranges (a measure of the equatorial variation of the radius) are large.  As an example, I show in Figure \ref{hektor} the rotational lightcurves \index{Lightcurves} of Trojan asteroid (624) Hektor at four different epochs.  The lightcurve range and shape change dramatically as the aspect angle ($\theta$, the angle between the line-of-sight and the pole) changes, but all the variations are well-modeled by an equilibrium Roche binary configuration \cite{cha87}, from which the density $\rho$ = 2480$_{-300}^{+80}$ kg m$^{-3}$ is deduced \cite{lj07}.   A check of this density is provided by the motion of Hektor's newly-found 15 km satellite \cite{mar06} and the assumption of Kepler's law.  The result, $\rho \sim$ 2200 kg m$^{-3}$ (Frank Marchis, private communication, August 2006), confirms the value found from the lightcurve model.  This fact, plus the remarkable quality of the fits in Figure \ref{hektor}, suggests that the shape of Hektor cannot be far from an equilibrium (strengthless) binary.  I speculate that impact jostling might explain why internal friction is unimportant: impacts energetic enough to cause bouncing or lifting of the components in an aggregate would allow the body to approach a near-equilibrium configuration by temporarily removing pressure-induced shear strength, just as strong vibrations cause a sand pile, initially at the angle of repose, to flow downhill against the inhibiting effects of inter-grain forces.  Whatever the cause, the lightcurves in Figure \ref{hektor} show that Hektor is well described as a strengthless equilibrium figure. \\

\begin{figure}
\centering
\includegraphics[height=7cm]{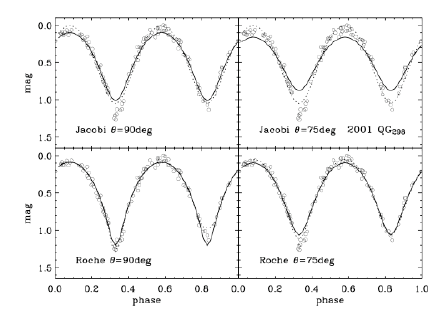}
%
%
\caption{Lightcurve of KBO 2001 QG298 compared with models. The top two panels show the best-fitting Jacobi ellipsoid models for aspect angles $\theta$ = 90$^{\circ}$ and $\theta$ = 75$^{\circ}$.  The bottom two panels show best fit Roche binary models for the same aspect angles.  The Roche binary model for 
$\theta$ = 90$^{\circ}$ (lower left panel) provides the best fit to the data, including the asymmetric lightcurve minima.  No comparably good Jacobi (single-body) models were found.  Data from \cite{she04}, figure from \cite{lj07}.}

\label{qg298}       
\end{figure}

\begin{figure}
\centering
\includegraphics[height=6cm]{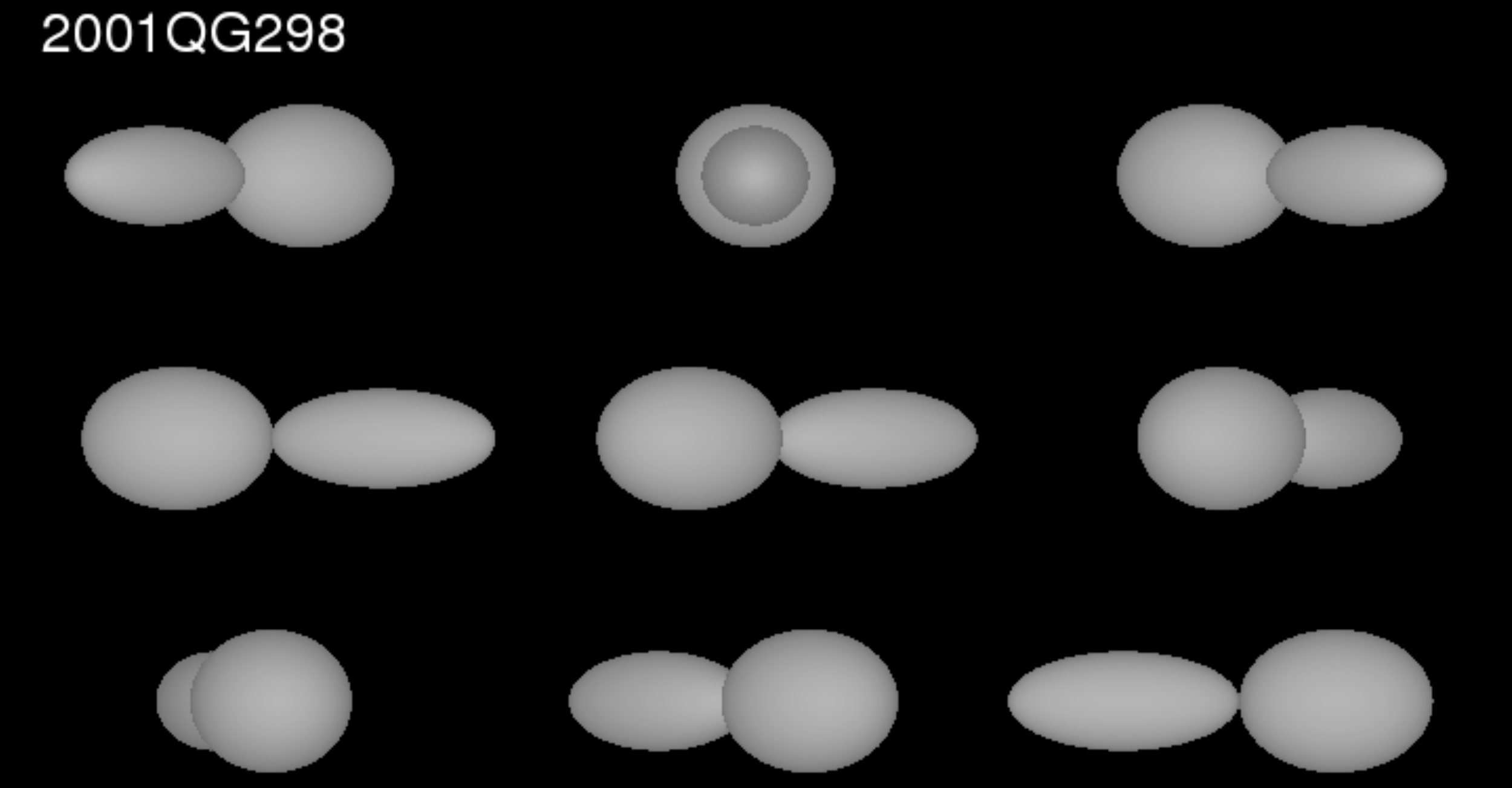}
%
%
\caption{Visualization of 2001 QG298 as a function of rotational phase based on the best-fit Roche binary model from the lower-left panel in 
Figure \ref{qg298}. The binary components are elongated by mutual gravitational attraction.  Figure from \cite{lj07}.}

\label{qg298_render}       
\end{figure}

Within the context of strengthless equilibrium models, we note that Jacobi ellipsoids generate lightcurves having a maximum range of $\sim$0.9 magnitudes \cite{wei80}, \cite{leo84}.  At more extreme rotations, the equilibrium configuration is a double object (a contact or near-contact binary\index{Contact binary}).  Therefore, objects with photometric ranges $>$0.9 mag., like Hektor itself (Figure \ref{hektor}), attract special attention as candidate contact binaries.  Several examples exist in the literature, including some in the main asteroid belt \cite{leo84}, the Kuiper belt \cite{she04} and there are others amongst the Trojans of Jupiter \cite{man07}, \cite{lj07}.  

Figure \ref{qg298} shows the mid-sized (effective diameter $\sim$ 240 km)  Kuiper belt object 2001 QG298 \cite{she04}, \cite{taka04}, \cite{lj07}.  Overplotted models confirm that the Jacobi ellipsoid models cannot fit, in particular, the deeply notched lightcurve minima.  The latter are better-fitted by Roche binary models where they are interpreted as mutual eclipse phenomena in a close binary (see Figure \ref{qg298_render}).  The density of 2001 QG298 given by a Roche binary fit to the lightcurve is $\rho$ = 590$_{-50}^{+140}$ kg m$^{-3}$ \cite{lj07}.  There is no \textit{proof} that 2001 QG298 is a Roche binary, but the ease with which the Roche binary model fits the lightcurve data suggests that this interpretation is plausible.  

Aside from the derived densities \index{Density}(discussed in more detail in the next section), the contact binaries may eventually help us to discriminate between various suggestions for the formation of binaries.  This is especially so in the Kuiper belt, where the fraction of binary objects is high \cite{ste06} and several formation mechanisms have been proposed.  Very briefly, these mechanisms include 1) binary formation in a debris ring created by a giant impact (as is thought to account for the formation of Earth's Moon and some large KBO satellites \cite{can05}) 2) permanent binding of a transient binary owing to the loss of energy by dynamical friction \cite{gol02} and 3) permanent binding via three-body reactions including exchange reactions \cite{fun04}.  All the proposed mechanisms require Kuiper belt number densities much higher than are now found, suggesting that binaries are products of a past epoch in which the Kuiper belt mass might have been substantially (by two to three orders of magnitude) higher than now.  

It is too early to reach any strong conclusion about the origin of the binary Kuiper belt objects.  For example, three-body interactions are weak and so produce mainly wide binaries, of which we know many examples \cite{ste06}.  Persistent drag from dynamical friction would cause steady inward spiraling of binaries, perhaps ending with the production of contact or very close binary systems (c.f. \cite{lj07} and Figure \ref{qg298}).   It seems likely that future determinations of the properties and statistics of the binaries, especially measurements of the contact to wide-binary ratio, will tell us a lot about the relative contributions to the binary population of different formation mechanisms.

\subsection{Densities}
\begin{figure}
\centering
\includegraphics[height=9cm]{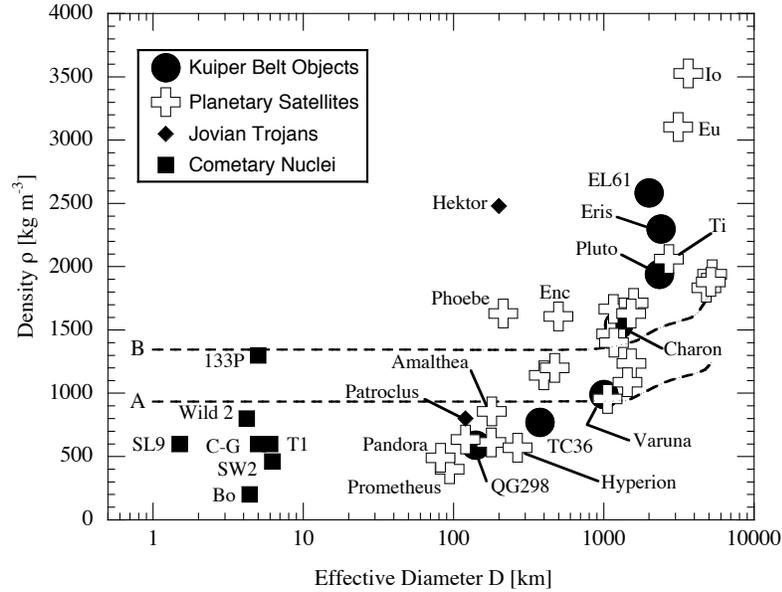}
%
%
\caption{Density as a function of diameter for mostly icy bodies in the outer Solar system.  Abbreviations SL9=D/Shoemaker-Levy 9, C-G=P/Churyumov-Gerasimenko, SW2=P/Schwassmann-Wachmann 2, Bo=P/Borrelly, T1=P/Tempel 1, QG298=2001 QG298, TC36=1999 TC36, EL61=2003 EL61, Enc=Enceladus, Ti=Titan, Eu=Europa.  Labeled curves are isothermal self-compression models for (A) pure water ice and (B) a 40\% rock and ice mixture from \cite{lup79}, for comparison purposes only (see text). Figure modified from \cite{jew07}.}

\label{density}       
\end{figure}

Figure \ref{density} (see also \cite{jew07}) shows the densities\index{Density} of objects in various small body populations as a function of the effective diameters.  Density data were obtained using a wide range of techniques, including gravitational perturbations on passing spacecraft (for the planetary satellites), mutual event data (for Pluto and Charon), the lightcurve models discussed above (for the other Kuiper belt objects\index{Kuiper belt objects}) and a mixture of (mostly) indirect techniques (for the cometary nuclei). 

While the range of densities at a given diameter is considerable, the tendency towards higher densities at larger sizes is self-evident in Figure \ref{density}.  There are no small bodies (diameters $D <$100 km) with high densities and no large bodies ($D >$ 1000 km) with densities much less than about 1000 kg m$^{-3}$.  The trend towards higher densities at larger sizes does not seem to be an artifact of mixing different samples having distinct sizes and densities.  For instance, the planetary satellites (hollow crosses in Figure \ref{density}) and the KBOs (large black circles) both show the trend toward densification as diameter increases in the range 100 km to 3000 km.

\begin{figure}
\centering
\includegraphics[height=10cm]{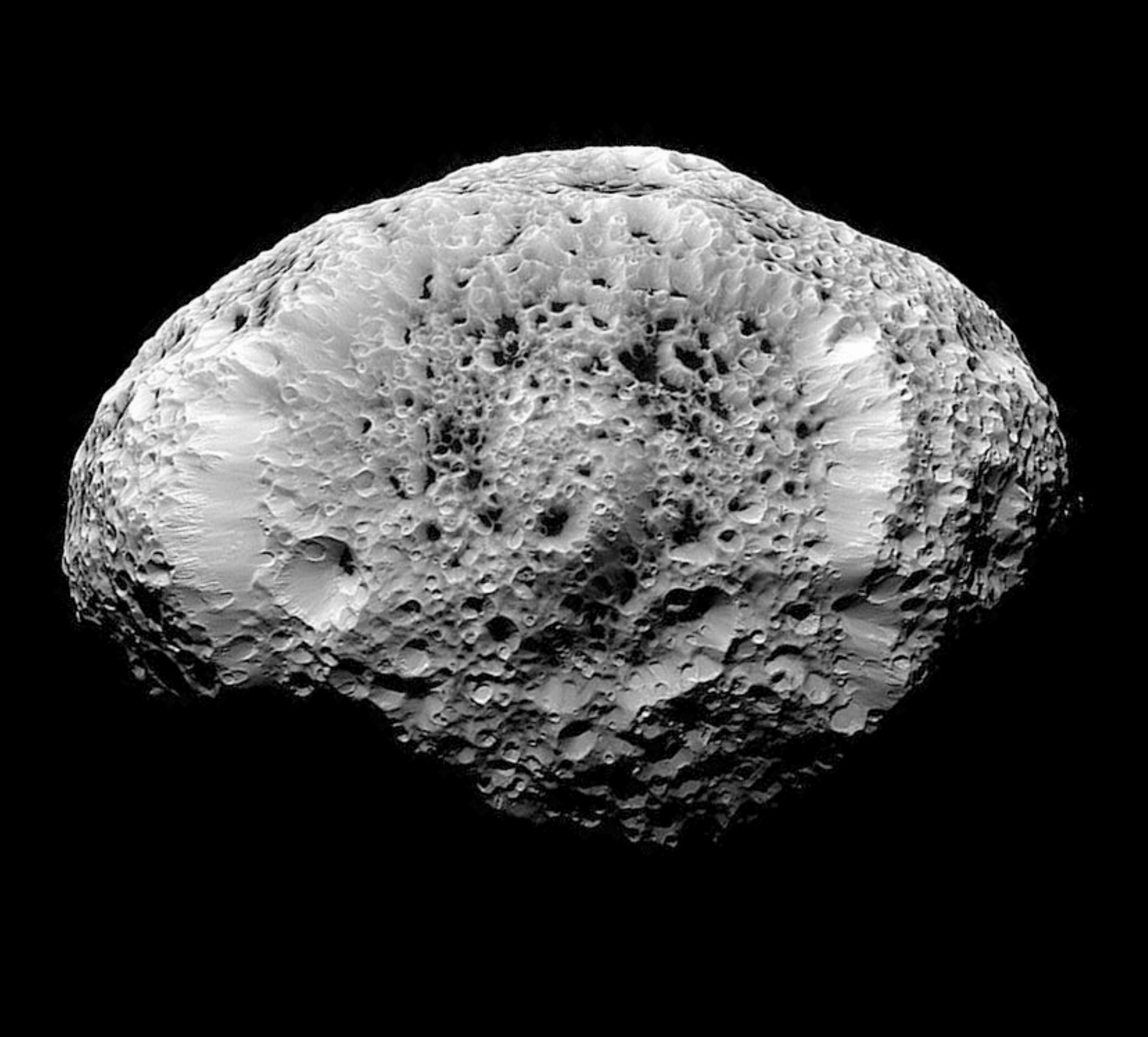}
%
%
\caption{Saturn's satellite Hyperion.  This aspherical body has a mean effective diameter of 270$\pm$8 km, a bulk density estimated from
perturbations on a passing spacecraft as $\rho$ = 540$\pm$50 kg m$^{-3}$ and a porosity \index{Porosity}$\sim$40\%   \cite{tho07}.  
Image courtesy Cassini Imaging Team and NASA/JPL/SSI.}

\label{hyperion}       
\end{figure}

The effects of self-compression on solid ice and rock-ice bodies are negligible for diameters $D<$1000 km, and modest even at the sizes of the largest objects plotted in Figure \ref{density}.  This is shown by the self-compressed models plotted as dashed lines in the Figure  (see \cite{lup79}). Some of the observed density vs. diameter variation must be compositional in origin.  For example, the large dense objects Io and Europa are largely rock-dominated while their similarly sized but less dense satellite companions Ganymede and Callisto have retained larger ice fractions.  On the other hand, compositional variations alone cannot account for objects with $\rho <$ 930 kg m$^{-3}$ (the density of uncompressed, pure ice \cite{lup79}).  Therefore, any object with a density less than 930 kg m$^{-3}$ \textit{must} be porous.
Clear examples of such low density, \textit{necessarily} porous objects \index{Porosity}are seen in Figure \ref{density} up to diameters of $\sim$500 km.  From Figure \ref{density} we see that the nuclei of most comets must be porous and Saturn's small satellites Pandora and Prometheus are so underdense that they also must be porous (a probable consequence of repeated collisional disruption and reassembly \cite{ren05}).  Another porous body is Jupiter's satellite Amalthea ($\rho$ = 860$\pm$100 kg m$^{-3}$, \cite{and05}), which was previously asserted to be one of the most refractory bodies in the Jupiter system but is now identified, amazingly, as a water-rich body \cite{tak04} more akin to a comet.  The Jovian Trojan (617) Patroclus ($\rho \sim$ 800 kg m$^{-3}$,  \cite{mar06b}), Kuiper belt contact binary 2001 QG298 (see above) and Saturn's tumbling moon Hyperion (see Figure \ref{hyperion}, $\rho \sim$ 540 kg m$^{-3}$, \cite{tho07}) all have low densities that require some fraction of internal void space even if they are composed of pure water ice.  Of course, it is hard to see how a pure water ice object could form.  Compositionally more realistic bodies with rock/ice ratios $\sim$1 would have $\rho \sim$1400 kg m$^{-3}$ or more.  Objects measured to have densities less than this value require some internal porosity.  For example, KBO (20000) Varuna, with $\rho \sim$ 1000 kg m$^{-3}$, must surely include both ice and rock and, depending on the exact rock/ice ratio and the nature of the rock component,  requires a porosity $\sim$20\% in order to explain the low density \cite{js02}, \cite{lj07}. 

The low densities could indicate microporosity (small internal voids with a scale comparable to the grain size) or macroporosity (internal void spaces with a larger scale) or some combination of the two.  Macroporosity might be generated by past collisional disruption followed by chaotic reassembly of the fragments.  This is a plausible explanation of the low densities ($\rho \sim$ 400 kg m$^{-3}$) of Saturn's strongly interacting co-orbital satellites Pandora and Prometheus, each about 100 km in diameter (see Figure \ref{density}).  The kilometer-scale nuclei of comets could also possess internal cavity space, since their gravitational self-compression is negligible.  However, I think it is unlikely that macroporosity is relevant in the deep interiors of very large, low density bodies (like Varuna \cite{js02}) where hydrostatic forces are appreciable (especially if these bodies have very low strength, as surmised from lightcurve data, above!).  The central hydrostatic pressure in a spherical object having diameter, $D$, and density, $\rho$, is $P_c \sim \pi$/6 $G \rho^2 D^2$ [N m$^{-2}$].  For example, with $D$ = 1000 km and $\rho$ = 1500 kg m$^{-3}$ the central pressure is $P_c \sim$ 7$\times$10$^7$ N m$^{-2}$. This is equivalent to the hydrostatic pressure 2.5 km below the Earth's surface, deeper than any known caves.  

Microporosity \index{Porosity} is a more likely candidate to explain the low measured bulk densities in Varuna-scale bodies and, if due to a loosely aggregated granular structure, would be more consistent with the low effective strengths of large bodies inferred from lightcurves.  Microporosity could be produced in the early Solar system as large bodies are assembled from smaller pieces, resting together much like grains of sand on the beach. (Incidentally, although it is not directly relevant to the case at hand, it is interesting to note that terrestrial beach sand is about 40\% porous at the surface and compresses to $\sim$25\% porosity at pressures of 0.5$\times$10$^8$ N m$^{-2}$, close to the core hydrostatic pressure on Varuna).   Laboratory experiments with compositionally relevant granular rock-ice mixtures show the evolution of microporosity in the 0.8 to 8$\times$10$^8$ N m$^{-2}$ pressure range \cite{leli94}, suggesting its potential importance for objects in Figure \ref{density}.  However, the temperature and its evolution through the life of the body will play an important role in determining the strengths of ice grains in outer Solar system bodies.  Therefore, it is necessary to compute coupled thermal-structural models to examine the long-term survival of porosity and this has barely been addressed \cite{leli94}.  Already, though, the data tell us that porosity must be significant in the outer regions of the 1000 km scale KBOs; at smaller sizes Figure \ref{density} shows that porosity can play a dominant role.\\

\textit{\textbf{Question:} To what degree do porosity variations and intrinsic compositional differences contribute to the different densities of objects of a given size in Figure \ref{density}?}

\textit{\textbf{Question:} To what extent are the porosities \index{Porosity}influenced by size-dependent thermal and ancient collisional processes?}

\section{Color Distributions}
\label{colors}

\subsection{Distribution of Colors}
\begin{figure}
\centering
\includegraphics[height=9cm]{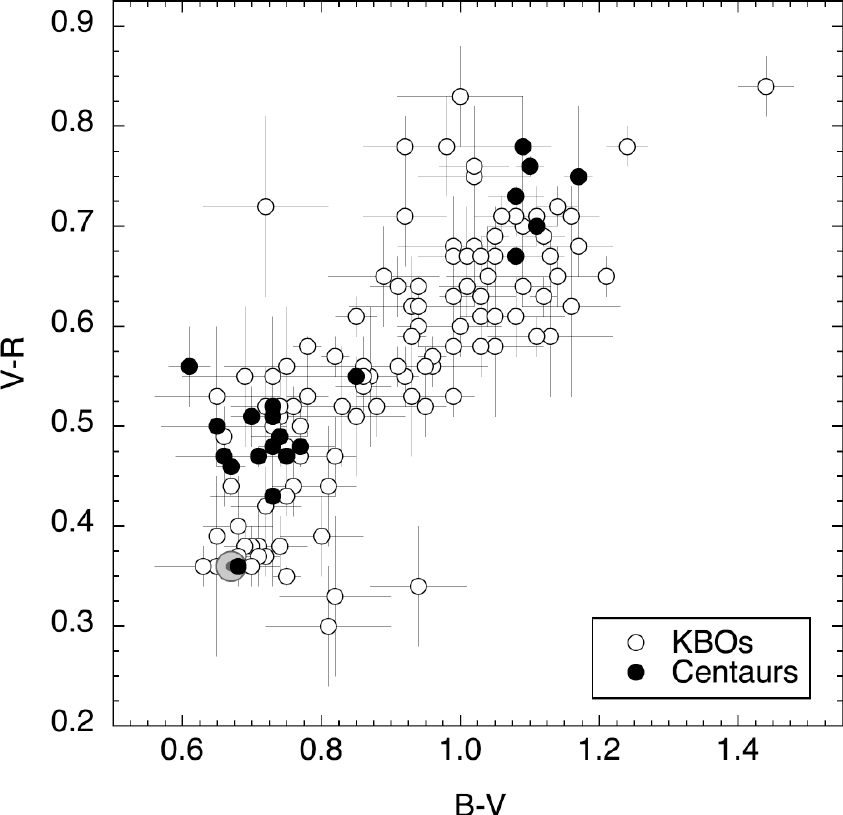}
%
%
\caption{$B-V$ vs. $V-R$ color-color diagram showing the KBOs (empty circles) and Centaurs (filled circles).  Only objects with 1$\sigma$ photometric uncertainties $<$ 0.1 mag. are plotted.  The Sun is marked by a grey circle.  Figure courtesy of Nuno Peixinho.}

\label{centaurkbo_bvr}       
\end{figure}

One of the first results to be established from systematic physical measurements of the Kuiper belt objects was that the
optical colors \index{Optical colors}are very diverse, ranging from approximately ``neutral'' ($V-R \sim$ 0.35) to ``very red'' ($V-R \sim$ 0.75) \cite{luu96}.  [$V$ and $R$ are the apparent magnitudes in filters centered near 5500\AA~and 6500\AA, respectively].  This 
finding was soon extended to the near-infrared, leading to the realization that the reflection characteristics of the KBOs are
determined over the wavelength range 0.45 $\le \lambda [\mu m] \le$ 1.2 by a single coloring agent \cite{jl01}, \cite{mcb03}.  This is different from the case of the main-belt asteroids where,
for example, distinct solid-state absorptions cause the spectral slope to vary dramatically with wavelength across this range.  There is widespread suspicion 
(but no compelling proof) that irradiated organics are responsible for the colors of at least some KBOs: such materials display the low
(few \%) albedos seen on many KBOs and can be very red (e.g. see \cite{mor03}).    The broad color dispersion has been confirmed by numerous independent measurements over the past decade.  This can be seen in Figure \ref{centaurkbo_bvr}, which is a compilation of published and on-line color measurements provided by Nuno Peixinho.  

Explanations for the color dispersion remain controversial.  In the resurfacing model (\cite{luu96}), the color of an object is set by competition between irradiation and impact-produced resurfacing.  Resurfacing excavates fresh material from beneath the surface layer susceptible to cosmic ray damage, thereby changing the surface color and (presumably) albedo.  
Observational evidence \textit{against} the resurfacing hypothesis is the lack of rotational variability of the surface colors: hemispheric color asymmetries caused by partial resurfacing should be more common than the data suggest \cite{jl01}.  Could intrinsic differences in the compositions of the KBOs cause the color dispersion?  Color differences in the main-belt asteroids are explained in this way but, in the Kuiper belt, compositional differences are less easy to understand.  Colors and compositions of main-belt asteroids are clearly related to the orbital parameters (especially semi-major axis) but similar correlations are not observed in the KBOs.   Further more, temperature differences between the inside of the Classical belt at $\sim$35 AU and the outside at $\sim$50 AU are only $\sim$10K, seemingly too small to have a major effect on the composition.   

Evidence for color-orbit correlations in the Kuiper belt is very limited. \index{Optical colors} 
An early claim \cite{teg98} that the optical colors of KBOs are distributed bimodally (i.e. that KBOs are \textit{either} neutral \textit{or} very red, but rarely in between) seems not to have survived independent scrutiny (Figure \ref{centaurkbo_bvr}).  Evidence that the colors of Centaurs are bimodally distributed is more convincing (\cite{pei03}; Figure \ref{centaurkbo_bvr}) but is unexplained.   The $B-R$ colors are related to perihelion distance \cite{teg00} or to the orbital inclination \cite{tru02}, but only for the Classical KBOs, a relation which is also unexplained.\\

\textit{\textbf{Question:}  What causes the color diversity on KBOs? }

\textit{\textbf{Question:}  Why are the Centaur colors bimodal?  In particular, if the Centaurs are escapees from the Kuiper belt, why do they not show the same colors and (unimodal) color distribution as the KBOs?  }

\textit{\textbf{Question:} Do the colors tell us something fundamental about the bulk compositions of these bodies, or do they merely reflect superficial processes acting on the optically accessible surface skin?}

%

\subsection{Ultrared Matter}\index{Ultrared matter}
The nearly linear reflectivity spectra of many outer Solar system bodies are usefully characterized by their gradients, 
expressed as $S'$ [\%/1000\AA] (\cite{jm86}).  Spectra with $S' >$ 25 \%/1000\AA~ are defined as ``ultrared'' \cite{jew02}.  Empirically, 
ultrared matter is found on the surfaces of Kuiper belt objects and Centaurs but is rare or absent on the surfaces of
small-bodies in other populations, including the Trojans \cite{dot06}, the cometary nuclei \cite{jew02}, dead JFCs \cite{jew02}, Damocloids \cite{jew05} and (perhaps) 
the irregular satellites (\cite{gra07}: however, too few of the latter have been adequately observed to be sure).
This lower incidence suggests that the ultrared matter maybe thermodynamically (or otherwise) unstable in bodies which approach
the Sun more closely than the Centaurs (which, by definition, have perihelia outside Jupiter's orbit).  

\section{Spectroscopy of Primitive Matter}
\label{spectra}
The wavelengths of vibrational and overtone spectral features of common molecular bonds fall into the
near infrared portion of the electromagnetic spectrum.  Accordingly, it is expected that near-infrared data should
place the most stringent constraints on the surface compositions of primitive Solar system bodies, both in the
inner and outer regions.  The faintness of many of the most interesting objects demands the use of large
telescopes, all of which are ground-based telescopes.   Nevertheless, the utility of near infrared spectra is limited by the faintness of the targets and by the difficulty of removing 
telluric signatures from the spectrum (the Earth's atmosphere contains many of the same molecular bonds
as those sought in the small bodies).  

Most objects studied in the near infrared show spectra which are utterly featureless.  

\subsection{Crystallinity of Solar System Ice}\index{Crystalline ice}
Ice can form at low temperatures in the amorphous \index{Amorphous ice}state, meaning that the geometric arrangement of the water
molecules lacks periodicity.  The amorphous state is distinct from the various crystalline forms in which water ice 
at higher temperatures is stable (e.g. the snow that falls from the sky and the ice that grows in the refrigerator is crystalline, with the
molecules arranged in staggered layers having a hexagonal pattern).  Amorphous ice is intrinsically unstable, and 
spontaneously transforms to crystalline ice on a timescale, $\tau_{cr}$ [yr], given by

\begin{equation}
\tau_{cr} = 3.0\times10^{-21} e^{\left[\frac{E_A}{kT}\right]}
\label{crystal}
\end{equation}

\noindent where $E_A$ is the activation energy, $k$ is Boltzmann's Constant, $T$ is the temperature and $E_A$/$k$ = 5370 K \cite{sch89}. 
The phase transition is potentially important for two reasons.  

First, the transition is exothermic, with a specific energy release 
$\Delta E$ = 9$\times$10$^4$ J kg$^{-1}$.  This $\Delta E$ can heat surrounding ice, influencing the thermal regime in icy bodies and perhaps even driving a runaway in which crystallization at one location in a body triggers crystallization over a large, thermally connected volume.  Crystallization is also associated with a small change in the bulk density.  Many elaborate and spectacular 
thermal models of comets are predicated on the assumption that the nuclei enter the middle and inner Solar system as amorphous ice bodies (\cite{pri04}).

Second, amorphous ice possesses many nooks and crannies, giving a large surface area per unit mass (of order 10$^2$ m$^2$ kg$^{-1}$ \cite{bar87}) on which other molecules can be trapped.    Empirically, a fit to experimental data (\cite{bar88}) on the trapping efficiency 
(defined as $\Re$ = $\frac{m_g}{m_i}$, where $m_g$ is the mass of gas that can be trapped in a mass of amorphous water ice, $m_i$) is given by

\begin{equation}
\Re \sim 10^{-0.08(T-40)}.
\label{barnun}
\end{equation}

\noindent Equation \ref{barnun}, which applies to $CH_4$, $CO$, $Ar$ and, to a lesser extent, $N_2$, gives $\Re \sim$ 1 at $T$ = 40 K, falling steeply to $\Re \sim$ 10$^{-5}$ at $T$ = 100 K.  At the $T \sim$ 40 K to 50 K temperatures prevalent in the Kuiper belt\index{Kuiper belt}, it is clear that large quantities of gas, 0.1 $\le \Re \le$ 1, could be trapped within amorphous ice, in agreement with observations of comets.  The trapped molecules are released as the temperature is raised above the accretion temperature, culminating with wholescale expulsion as the water molecules rearrange themselves into cubic or hexagonal lattices upon crystallization. The presence of amorphous ice can thus lead to pulses of outgassing that could be relevant to understanding the mass loss from comets.  

\begin{figure}
\centering
\includegraphics[height=8cm]{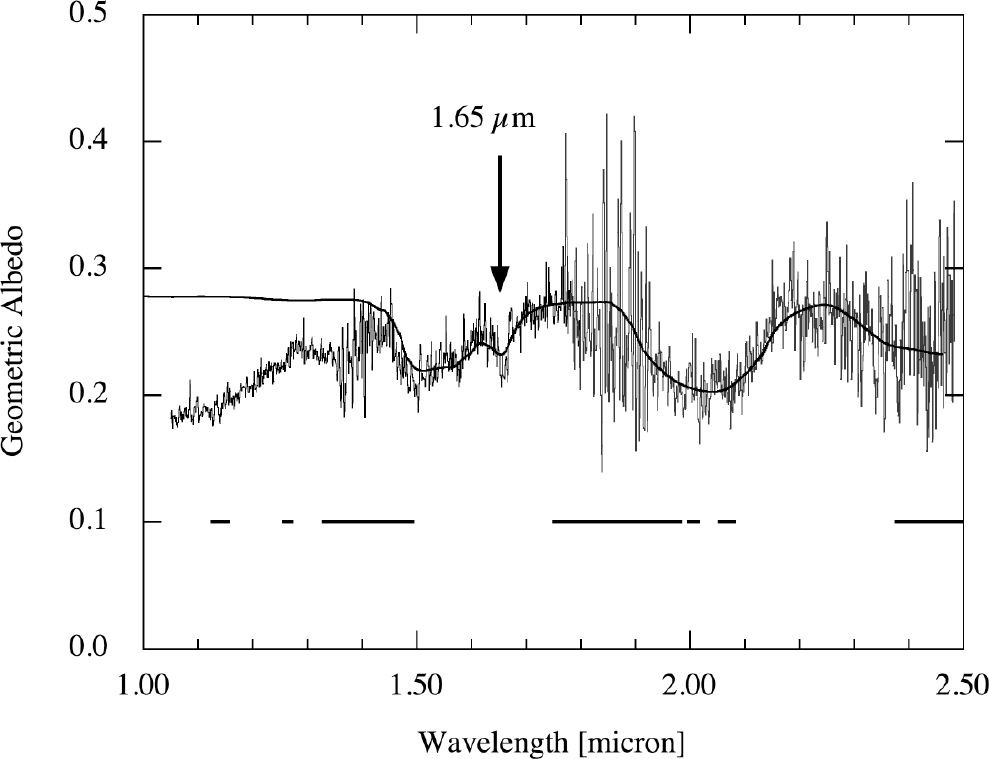}
%
%
\caption{Near infrared spectrum of Kuiper Belt Object (50000) Quaoar showing the
major ice bands at 1.5 $\mu$m and 2.0 $\mu$m and the narrow feature at 1.65 $\mu$m
that is diagnostic of the presence of crystalline ice\index{Crystalline ice}.  The smooth line is a crystalline ice spectrum that has been plotted on top of the Quaoar spectrum for comparison: no attempt was made to fit the data but still the correspondence between Quaoar and the ice spectrum is impressive. Horizontal bands at the bottom of the figure show regions where the transparency of the Earth's atmosphere is particularly poor.  From \cite{jl04}.}
\label{quaoar}       
\end{figure}

By setting $\tau_{cr}$ = 4.5$\times$10$^9$ yr in Equation \ref{crystal}, we find that amorphous ice formed at the beginning of the Solar system would have escaped crystallization if its temperature had always been $T <$ 77 K.  Because of the very strong temperature dependence in Equation \ref{crystal}, even a brief excursion above this temperature would have crystallized the ice.  The temperature of an isothermal blackbody in thermal equilibrium with sunlight falls to 77 K at $R$ = 13 AU, or slightly beyond the orbit of Saturn.  Therefore, all else being equal, we should expect to find crystalline ice at (and inside) the orbit of Saturn, and to find amorphous ice beyond.  Water ice in the inner regions is indeed crystalline, but  it is also crystalline in the satellites of Uranus and Neptune and in the Kuiper belt.  There is surprisingly no direct evidence for amorphous ice in the outer regions (see Figure \ref{quaoar}).  

The two types of ice are observationally separable in the near infrared.  The 1.5 $\mu$m and 2.0 $\mu$m bands have slightly different shapes and central wavelengths, but a much better diagnostic is provided by the crystalline ice band at 1.65 $\mu$m.  This band is absent in amorphous ice.  If the 1.65 $\mu$m band is present then the ice must be at least partly crystalline.  If it is absent then the ice might be amorphous, down to some limit set by the signal-to-noise ratio of the spectrum around the band.

However, the optically observable surfaces of bodies are bombarded
by energetic particles from the Solar wind and from cosmic rays\index{Cosmic ray irradiation}, and also by energetic photons from the Sun.  These energetic particles disrupt the bonds between water molecules in ice, thereby breaking up the crystal structure and ``amorphizing'' the material. (It is interesting to note that silicate grains in the interstellar medium are largely amorphous for the same reason \cite{kem04}).  The timescale for amorphization is short, probably 10$^6$ yr to 10$^7$ yr \cite{jl04}.  In this sense, the presence of crystalline ice in the outer Solar system is even more surprising and the reason for its persistence has not yet been firmly explained.  One possibility is that resurfacing provides fresh material on a timescale that is short compared to the amorphization time.  Resurfacing could result, for example, from impact gardening, which dredges up buried material (ice deeper than $\sim$1 meter is effectively shielded from even quite energetic cosmic rays).
A more dramatic possibility is that outgassing or cryovolcanism emplaces fresh, crystalline ice on the surface.  Very recent work with an ultra-high vacuum chamber in the Chemistry Laboratory of the University of Hawaii suggests a more likely explanation. We find that the amorphization efficiency is a function of temperature such that amorphization is nearly 100\% efficient at $T \sim$ 10 K but only $\sim$50\% efficient at $T$ = 50 K.  Presumably, this is because slight thermal jostling at the higher temperatures allows some water molecules to reconnect in the crystalline form even after irradiation \cite{zhe08}.  At the surface temperature of Quaoar (Figure \ref{quaoar}), ice can remain partly crystallized forever, despite the rain of energetic particles.

While the persistence of crystalline ice is apparently now understood, what heated the ice to make it crystalline in the first place remains unknown.  Several possibilities exist.  In large bodies (radii $>$500km) it is possible that heating occurred upon formation by the conversion of gravitational potential energy into heat.  Large bodies could also have been heated by trapped radionuclides, whether they be short-lived (half-lives $\sim$10$^6$ yr) like the famous $^{26}$Al and $^{60}$Fe, or long-lived (half-lives $\sim$10$^9$ to 10$^{10}$ yr) like $^{40}$K, $^{232}$Th and $^{238}$U.  Local surface heating by micrometeorite bombardment has the advantage that it would operate on bodies of any size, consistent with crystalline ice being common in the outer Solar system on objects of different diameters.  Whatever the cause, the available evidence shows that ice on the surfaces of the large Kuiper belt objects is crystalline, which means that it has been warmed at least to twice the current surface temperatures of 40 K or 50 K.

\begin{figure}
\centering
\includegraphics[height=9cm]{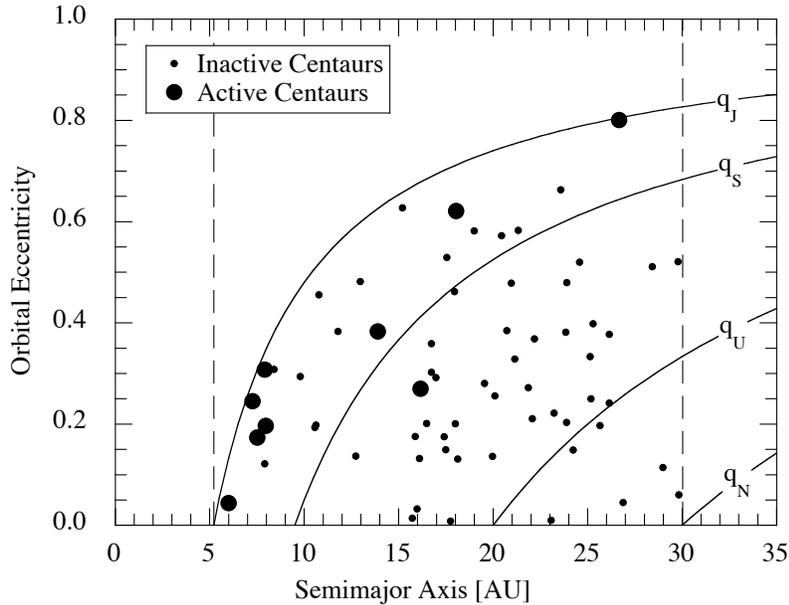}
%
%
\caption{Distribution of the Centaurs in semimajor axis vs. eccentricity space.  Large circles denote active (outgassing) Centaurs while small circles show inactive Centaurs.  The semimajor axes of Jupiter and Neptune, which bound the Centaur orbits, are shown with vertical dashed lines.  
Diagonal arcs show the loci of points having a fixed perihelion distances equal to the semimajor axes of the orbits of the giant planets, as marked.   \cite{j07b}.}
\label{centaurs}       
\end{figure}

Small bodies, like the nuclei of comets, were probably not substantially heated by the above processes.  Do they contain amorphous ice\index{Amorphous ice}?  Only limited direct evidence exists in the form of spectra of the dust in two long-period comets, both distinguished by showing no evidence for the 1.65 $\mu$m crystalline ice band.  Other evidence comes from the distribution of the orbits of the Centaurs\index{Centaurs}.  These are objects recently escaped from the Kuiper belt and traveling on orbits which cross the paths of the giant planets (i.e. their defining property is that they have perihelia and semimajor axes between the orbits of Jupiter and Neptune).  About 20\% of the known Centaurs are also active comets.  The distribution of the orbital elements of the active Centaurs is different from the Centaurs as a whole.  In particular, the average perihelion distance of the active Centaurs is small compared to the average perihelion of the Centaurs as a whole.  This difference cannot be ascribed to the simple sublimation of crystalline water ice, since the latter is involatile throughout the Centaur region.  Instead, activity in the Centaurs is consistent with production through the crystallization of amorphous ice, which begins at temperatures comparable to those found on the active Centaurs when at perihelion \cite{j07b}.  This is not iron-clad evidence for the existence of amorphous ice in the Centaurs, by any means.  But it is perhaps the best evidence we possess at the moment. \\

\textit{\textbf{Question:}  Can more objects be observed in order to determine whether the ice is truly crystalline in these objects?  Spectra of adequate quality have been secured for only two comets.   Are only the long-period comets amorphous?  What about Halley-family comets?}

\textit{\textbf{Question:}  What crystallizes ice on the larger Kuiper belt objects and other bodies in the outer Solar system?  Is it a global energy phenomenon as
suggested (e.g. gravitational binding energy, or decay of trapped radoiactive nuclei) or merely a surface effect (e.g. micrometeorite heating and crystallization of a thin surface layer)?}

\subsection{The Methanoids}
\begin{figure}
\centering
\includegraphics[height=9cm]{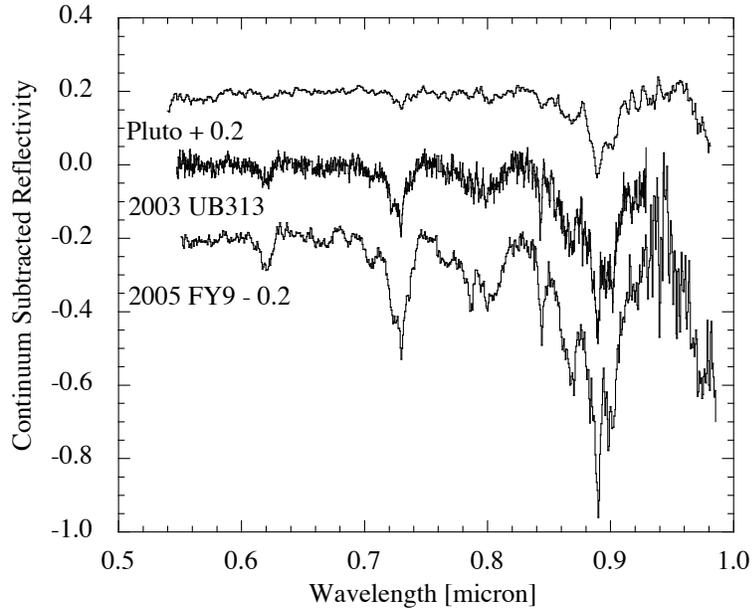}
%
%
\caption{Far-red optical spectra of the three methanoids\index{Methanoids} (134340) Pluto, (136199) Eris (formerly 2003 UB313) and (136472) 2005 FY9, taken at the Keck 10-m telescope.  The spectra are continuum-subtracted and vertically displaced for clarity.  All the visible absorption bands in these spectra are due to solid methane.}

\label{methanoids}       
\end{figure}

\index{Methanoids}
Water ice is present on some large KBOs while others show instead prominent bands due to methane \cite{lic06}, \cite{teg07}.
These ``methanoids'' include amongst their number (134340) Pluto, as well as (136199) Eris and (136472) 2005 FY9 (Figure \ref{methanoids}).  Jeans (thermal) escape appears to determine which KBOs can retain $CH_4$ and which cannot: methane is more stable on the large, distant (cold) KBOs than on small, close (hotter) ones \cite{sch07}.  

The source of the methane is unknown.  One possibility is that the methane is produced, along with other hydrocarbons, as a by-product of energetic particle irradiation of exposed surface ices. A pre-existing source of carbon would need to be present within the ice in order for $CH_4$ to be formed this way.  In this case, one might expect all large and cold KBOs to show methane, since all are comparably irradiated by the solar wind and cosmic rays.   Alternatively, perhaps methane was delivered to the KBOs at the time of their accretion in the form of clathrated ice (but this might be difficult to reconcile with the picture outlined above in which low temperature ice making up the KBOs is more likely to have been amorphous, at least at the accretion epoch).  The most exciting possibility is that the methane has been created through chemical reactions in the deep interiors of the larger KBOs and has since leaked onto the surface.  We know from Terrestrial experience that many serpentinization reactions (between liquid water and rocks) are exothermic and release hydrogen (\cite{fru03}).  Fischer-Tropsch type reactions between the hydrogen so-produced and carbon monoxide could create methane.   The main requirements for the active generation of methane would then be the existence of liquid water significantly above the triple point and intimate contact with carbon-containing rocks over a large reaction surface.  Both circumstances appear likely in the larger (1000 km scale) KBOs \cite{bus03}, \cite{mer06}.   

Lastly, it is good to keep in mind that while Nature always plays by the rules, it doesn't always play fair: it is entirely possible that more than one source contributes $CH_4$ to the methanoids and equally likely that the dominant source is not one that we have thought of. \\

\textit{\textbf{Question:} How can we decide between alternative production schemes for methane, and what others might exist?}

\textit{\textbf{Question:} How could internally generated methane move from the deep interior of a KBO to the surface? Which other volatiles would move with it?}

\textit{\textbf{Question:} Can we detect atmospheres of KBOs other than Pluto, perhaps by the occultation of background stars?}

\section{Irregular Satellites}\index{Irregular satellites}
\label{isats}
For the most part, the satellites of the planets can be neatly separated into one of two distinct categories based on their orbits.
The so-called regular satellites \index{Regular satellites}have small orbital inclinations and eccentricities ($e \ll$ 1).  By contrast, the irregular satellites (hereafter ``iSats'') have large inclinations
(spanning the range 0 $\le i \le$ 360$^{\circ}$: most irregulars are retrograde) and eccentricities ($e \sim$ 0.5).  Another distinction is based on the fraction of the Hill sphere occupied by the
orbits of the satellites.  The Hill sphere\index{Hill sphere} is the volume in which a planet exerts gravitational control of nearby objects in competition with the Sun.  The Hill sphere radius is $r_H = a [m_p/(3 M_{\odot}]^{1/3}$, where $m_p/M_{\odot}$ is the mass of the planet in units of the Solar mass and $a$ is the semimajor axis of the orbit of the planet.  [Values of $r_H$ are given in Table \ref{hillspheres} both in AU and in apparent angle on the sky as seen from Earth.  The Table also lists the (ever changing) numbers of known satellites at each planet].  A general rule is that orbits of the regular satellites are confined to the central few percent of $r_H$ while most iSats are much more wide-ranging, with orbital semimajor axes up to $\sim$0.5 $r_H$.  Although their orbits, and the effects of Solar tides, are very large, the known iSats appear to remain bound to their planets for timescales comparable to the age of the Solar system.

\begin{table}
\centering
\caption{Hill Spheres of the Giant Planets (from \cite{jh07}) \label{hillspheres}}\index{Hill sphere}
\begin{tabular}{lcccccc}
\hline\noalign{\smallskip}
Planet & Mass [M$_{\oplus}$] & $a$[AU] & r$_H$ [AU] & r$_H$ [deg] & N$_r$ & N$_i$ \\
\hline\noalign{\smallskip}
Jupiter   & 310  &  5 & 0.35 & 5 & 8 & 55 \\
Saturn   & 95  &  10 & 0.43 & 2.8 & 21 & 35\\
Uranus   & 15  &  20 & 0.47 & 1.4 & 18 & 9\\
Neptune   & 17  &  30 & 0.77 & 1.5 & 6 & 7\\
\noalign{\smallskip}\hline
\end{tabular}

NOTE: N$_r$ (N$_i$) are the numbers of regular \\
(irregular) satellites at each planet. 
\end{table}

These systematic differences in the orbital inclinations, eccentricities and sizes (relative to $r_H$) reflect different modes of formation of the regular
and irregular satellites.  Whereas the regular satellites are clearly the products
of accretion in long-gone circumplanetary disks, the irregulars more likely formed in orbit about the Sun (but we don't know where) and were subsequently captured by
the planets (we like to know when and how).  

Most of the very large (i.e. bright) satellites fall in the ``regular'' class and, for this
reason, the regulars have captured most of our attention since Galileo discovered his four large (regular) satellites of Jupiter in 1610.  Recent observational work has refocused our attention by establishing that iSats substantially
out-number the known regular satellites and that the two types formed differently \cite{jh07}.  Irregular satellites have unambiguously emerged as a  ``hot topic'' in planetary science.

\begin{figure}
\centering
\includegraphics[height=9cm]{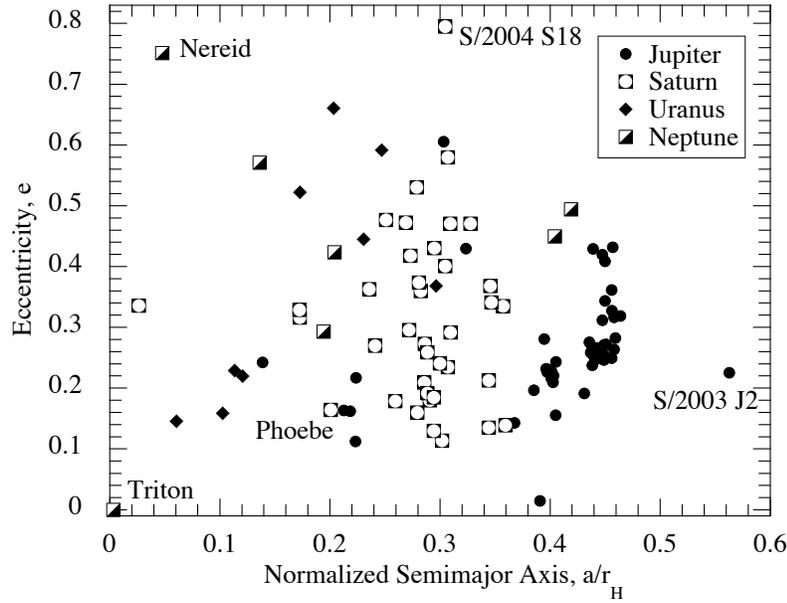}
%
%
\caption{Distribution of the irregular satellites in semimajor axis vs. eccentricity space.  Selected objects and dynamical groups are identified.  
Figure updated (to 2007 May) and adapted from \cite{jh07}.}
\label{mbcae}       
\end{figure}

There are several ideas about the origin of the iSats.  \index{Irregular satellites}Until recently, the most popular idea was that the satellites were
captured from heliocentric into planetocentric orbits through the action of gas drag\index{Capture!Gas drag}, in the extended atmospheres of the growing giant planets.  This idea was first proposed to account for the iSats of gas giant planet Jupiter \cite{pol79}.  It relies on the collapse of a massive gaseous envelope to provide a transient source of drag since, if the drag persists, all satellites must ultimately spiral down into the planet.    The idea might also work for the other gas giant, Saturn, but it is not so obvious that it can be applied to Uranus and Neptune, since these planets are \textit{ice} giants.  The ice giants have comparatively modest gas inventories (e.g. a few $M_{\oplus}$ compared with $\sim$80$M_{\oplus}$ and 260$M_{\oplus}$ in Saturn and Jupiter, respectively).  Moreover, the timescales of formation are completely different, probably $\sim$1 Myr or less for Jupiter and Saturn but 10 or more
times longer at Uranus and Neptune.  

\begin{figure}
\centering
\includegraphics[height=9cm]{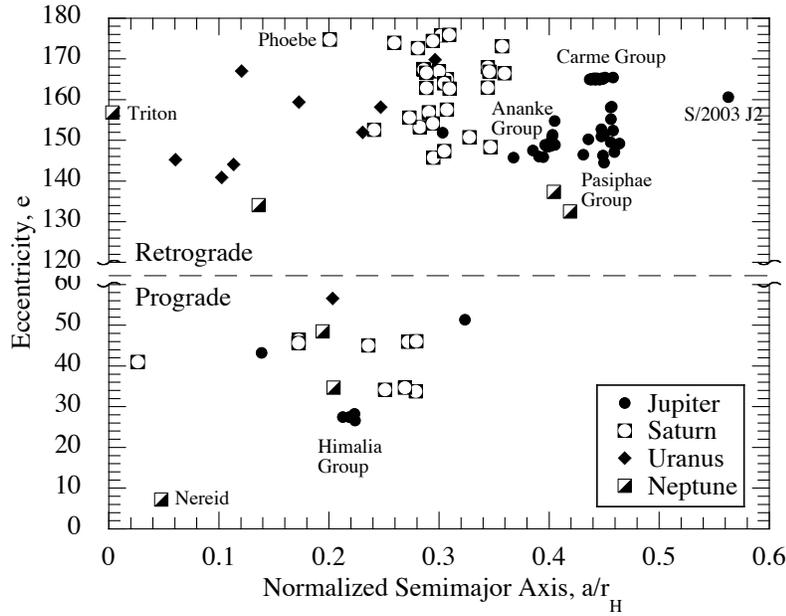}
%
%
\caption{Distribution of the irregular satellites in semimajor axis vs. inclination space. No known satellites have inclinations 60 $\le i \le$ 120$^{\circ}$ and so this region is not plotted.   Orbits in this range are unstable to the Kozai resonance.  Selected objects and dynamical groups are identified.  
Figure updated (to 2007 May) and adapted from \cite{jh07}.}
\label{mbcae}       
\end{figure}

A second idea is that the satellites were captured in a phase of runaway growth, when the gas giants were pulling in gas from the adjacent
protoplanetary disk.  Sudden growth in mass leads to sudden expansion of the region around each planet in which the gravitational 
influence of the planet dominates that of the Sun \cite{hep77}.  In this ``pull-down'' model\index{Capture!Pull down}, the iSats would have been captured objects that happened to be nearby to the planets at the end phases of their runaway growth.   One, apparently fatal, problem for this model is that the ice giant planets did not undergo runaway growth.  They accreted mass by binary collisions of solid objects over a long period of time (evidently comparable to or longer than the $\sim$10 Myr timescale on which gas survived in the disk), with steady growth but no mass runaway.  

The last idea has emerged as the most interesting, given what we now know about the young Solar system.  The idea is that irregular
satellites were captured from heliocentric orbits in three-body (or N-body) interactions \cite{col71}. \index{Capture!3-Body}  For example, the three bodies could be two planets and a small-body initially in orbit about the Sun (\cite{nes07}) or two asteroids could interact with each other within the Hill sphere of a planet \cite{col71}.  As a result of the interaction, one of the small bodies could be ejected from the planetary region, carrying with it excess energy that would allow the other asteroid to become bound.  One attraction of 3-body and N-body capture models is that the Hill spheres of the four giant planets increase in size and volume with increasing distance from the Sun (even though the masses of the giants decrease from Jupiter outwards).  One consequence might be that low mass, distant Uranus and Neptune might be able to capture about as many irregulars as high mass Jupiter and Saturn, in accordance with the data \cite{js05}.  However, this conjecture has not yet been placed on a quantitative basis.  Indeed, 3-body and N-body capture models have received scant attention probably because, until recently, it seemed that such interactions in the Solar system must be incredibly rare.  In the modern system such interactions \textit{are} rare, but they may not always have been so, since the early Solar system was much more densely populated than it is now.  

From where were the iSats captured?  \index{Irregular satellites}The evidence does not provide an answer to this question, so we remain for now in a state of conjecture.  The first main possibility is that the iSats were captured from initial heliocentric orbits that were close to, or at least crossing, the orbits of the giant planets.  Low velocity encounters give the highest probability of capture, so local sources are in some sense preferred.  The second possibility is that the iSats were captured from a remote source, perhaps the Kuiper belt.  The latter possibility has been advanced in the context of the ``Nice'' dynamical model \cite{gom05}, \index{Nice model}in which the architecture of the Solar system is a consequence of an assumed crossing of the 2:1 mean-motion resonance between Jupiter and Saturn.  According to initial simulations with this model, capture of the iSats of Uranus and Neptune (and perhaps Saturn) is possible but the iSats of Jupiter must have another source \cite{nes07}.  \\

\textit{\textbf{Question:} How and when were the iSats captured? Was there a single capture mechanism or did different planets capture their satellites in different ways?  How can we tell?}

\textit{\textbf{Question:} From where were they captured? From the Kuiper belt, from orbits in the protoplanetary disk, local to the growing planets, or from elsewhere?}

\textit{\textbf{Question:} Does ultrared matter \index{Ultrared matter}exist on iSats?  If the iSats were captured from the Kuiper belt, the presence or absence of ultrared matter might constrain the source region.}

\textit{\textbf{Question:} How do the answers to these questions change from planet to planet?}

\section{Main Belt Comets}
\label{mbc}

\begin{figure}
\centering
\includegraphics[height=7cm]{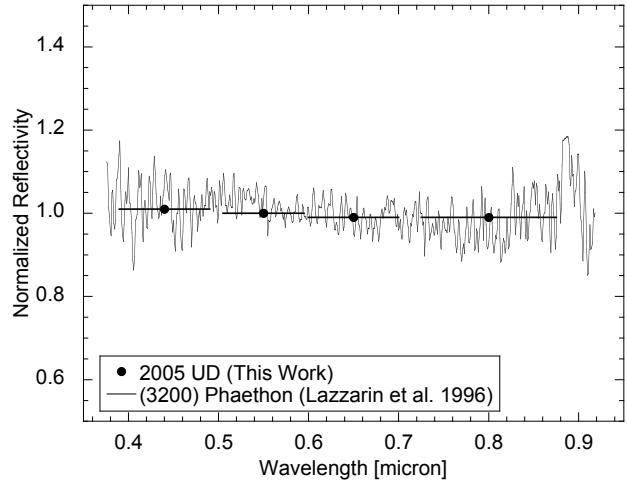}
%
%
\caption{Image of main belt comet 133P/Elst-Pizarro taken at the University of Hawaii 2.2-m telescope on UT 2007 June 11.  A dust tail is visible 
extending to the right of the nucleus.  The region shown is
approximately 70 arcsec in width and has North to the top, East to the left.  The MBC has approximate apparent red magnitude 19.5.}
\label{133Pb}       
\end{figure}

Main belt comets (MBCs) \index{Main-belt comets}are objects with orbits in the region classically occupied by the asteroids but with
physical characteristics of comets, specifically including comae and/or tails (Figure \ref{133Pb}).  Three examples are known as of July 2007 \cite{hsi06}.  Their Tisserand parameters \index{Tisserand parameter}measured
with respect to Jupiter are $T_J >$ 3, whereas those of comets from the Kuiper belt \index{Kuiper belt}and Oort cloud \index{Oort cloud}reservoirs are $T_J <$ 3.  The MBCs are also completely distinct from the more familiar ``transition objects'' (see Section \ref{transition}).  The latter, in fact, are the \textit{opposites} of the MBCs in having comet-like orbits (with $T_J <$ 3) but asteroid-like
physical appearances (i.e. no comae and no tails).  This difference is clear in Figure \ref{mbcae} and in the classification diagram in Figure \ref{schema}.

Evidence that the mass loss from MBCs is driven by the sublimation of ice is indirect.  Specifically, none of the other
mechanisms that we have thought of seem to fit the data.  The first suggestion for 133P was that the mass loss is impact debris, resulting from a small collision \cite{tot00}.  This explanation is now ruled out, given that the activity in 133P is periodic, having been present near perihelion in 1996, 2002 \cite{hsi04} \cite{tot06} and now again in 
2007 (Figure \ref{133Pb}).  Rotational instability seems an unlikely explanation.  While 133P
is rotating quickly (period = 3.47 hrs), there is no evidence for rapid rotation in either P/Read or 176P.  Moreover, there are
many asteroids rotating with shorter periods, yet these are not known to be emitting dust like the MBCs.  On the Moon, charge
gradients in the vicinity of the terminator are known to levitate and launch dust particles from the surface \cite{lee96}.  \index{Electrostatic charging} The same 
process could eject dust from small, low escape-velocity asteroids and comets.  Two problems with this mechanism for the MBCs are 1) that dust velocities inferred from 133P and P/Read
are higher than typical on the Moon and, more seriously, 2) if electrostatic ejection were important, we would have to ask why comet-like emission is not a general property of all small asteroids.   There is also an issue with supply.  Unlike the Lunar case, a large fraction of the small dust grains on asteroids are simply lost into space, not levitated repeatedly as the terminator sweeps by.  New dust particles will be created by micrometeorite impact into the asteroid surface, but the rate of production is orders of magnitude too low to account for the escape losses to space.

The MBCs \index{Main-belt comets}hold special significance in planetary science because they appear to be repositories of ice in a region of the Solar system
that has been suggested, on independent grounds, as a potential contributor to the Earth's oceans \cite{mor00}.   The reasoning behind this is as follows.  The Earth probably formed too hot to have accreted \textit{and} retained much water and so this, and other, volatiles were accreted from another source in a ``late veneer'' some time after the Earth had cooled down.  The timing of the addition of water is uncertain.  However, evidence from $^{18}$O isotopes in some zircons (ancient refractory mineral grains which substantially predate the rocks in which they are found) suggests that substantial bodies of liquid water were present very early, at 4.3 Gyr \cite{moj01} or even 4.404$\pm$0.008 Gyr \cite{wil01} ago.

\begin{figure}
\centering
\includegraphics[height=10cm]{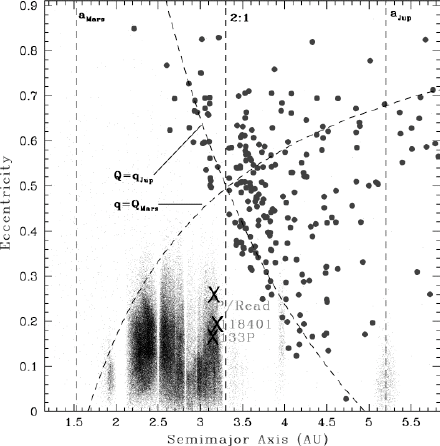}
%
%
\caption{Semimajor axis vs. orbital eccentricity for asteroids (small dots), Jupiter family comets (large dots) and the
three currently known main belt comets (marked X).  The latter are clearly associated more with the asteroid belt than with the
Jupiter family comets.  The semimajor axes of Mars and Jupiter are marked with vertical dashed lines.  Two labeled arcs show the
locus of orbits having perihelion inside Mars's aphelion distance and aphelion outside Jupiter's perihelion distance, respectively.
Figure from \cite{hsi06}.}
\label{mbcae}       
\end{figure}

Comets, being ice-rich, are one possible source of terrestrial water.  Against this are measurements showing that the $D/H$ \index{D/H ratio} ratios in comets are twice the $D/H$ ratio measured in the Earth's oceans.  Either the terrestrial $D/H$ has evolved (possible), or the cometary $D/H$ values are wrong (unlikely, see \cite{mei99}) or unrepresentative (possible, because the measured comets are not the Jupiter family comets most likely to have contributed water \cite{del98}) or the comets are not the dominant source of Earth's water \cite{mor00}.  The mass of the oceans is about 2.5$\times$10$^{-4} M_{\oplus}$.  The mass of water trapped within the mantle is very uncertain and could be much less than or much greater than the mass on the surface.  Dynamical models suggest that such a large mass is unlikely to have been trapped from the Kuiper belt and point instead to a closer source in the asteroid belt \cite{mor00}.  In this latter interpretation, the MBCs occupy a region that might have contributed to the oceans.  It is important to note that the objects now present in the outer belt cannot be suppliers of Earth's water: there are too few and there is no clear dynamical pathway from most of the outer asteroid belt to Earth-intersecting orbits.  What is imagined is that a massive, primordial asteroid belt was cleared (probably by strong perturbations from nearby Jupiter) at some earlier time, hurling ice-rich objects across the paths of the terrestrial planets.  

How could ice become trapped in the main belt asteroids?  On the surface there would seem to be two possibilities.  Either the ice originated there, becoming trapped in the MBCs as they formed, or the ice was delivered after formation from a more remote source.  The presence of hydrated minerals in many meteorites thought to come from the outer belt requires the past presence of liquid water (e.g. \cite{bre06}, \cite{jcgp07}).  Perhaps the MBCs are icy asteroids in which some of the primordial ice component escaped chemical reaction with silicates and persists to the present day.  I know of no evidence against this possibility.  On the other hand, attempts to capture comets from the Jupiter family into orbits like the MBCs seem doomed to fail.  The Tisserand parameter is approximately a constant of the motion during capture, and the fact that the MBCs and JFCs have different Tisserands indicates that simple conversion of the orbits is impossible.  Additional forces, from non-gravitational accelerations due to anisotropic outgassing or from perturbations by terrestrial planets, could conceivably help transform JFC orbits into MBC orbits.  I am open to this possibility and would like to see more work done to explore it.  What has been published on this topic, however, gives little reason to be optimistic \cite{lev06}.  

How can ice be stable in the main belt only $\sim$3.2 AU from the Sun?   The temperature of an isothermal blackbody located at this distance is $T_{BB}$ = 153 K.  $T_{BB}$ gives a good estimate of the averaged, deep temperature in kilometer sized MBCs, while regions on the surface, for example near the subsolar point, can be expected to be hotter.  The specific sublimation rate in thermal equilibrium at $T_{BB}$ is $dm$/$dt \sim$ 3$\times$10$^{-8}$ kg m$^{-2}$ s$^{-1}$.  An MBC surface having density $\rho$ = 2000 kg m$^{-3}$ would recede at the rate $\rho^{-1}$ $dm$/$dt \sim$ 1.5$\times$10$^{-11}$ m s$^{-1}$, corresponding to about 0.5 mm yr$^{-1}$.  A 1000 m radius body could survive for only $\sim$2 Myr, if in continuous sublimation at this rate, which is very short compared to the age of the Solar system.  Therefore, the ice must be stabilized against sublimation losses if it is to have survived for the age of the Solar system.  

Observations show that the nuclei of comets are mantled by refractory matter.  By analogy it seems reasonable to suppose that mantles also exist on the MBCs and that they stifle the gas flow from most or all of the surface, most of the time.  In this way, ice might survive in the MBCs for the age of the Solar system, even at distances considerably smaller than 3 AU.  Ice stability, protected by porous, refractory mantles, has been established for asteroid (1) Ceres at 2.7 AU \cite{fan89} and even for Mars' satellite Phobos at 1.6 AU \cite{fan90}.   In order to become visibly active, the mantle of an MBC must be punctured.  A likely mechanism in the main belt is collision.  A meter-scale impactor would expose enough ice to drive the mass loss rates that are inferred for 133P, for example.  

So, a plausible scenario for the MBCs \index{Main-belt comets}is that they are ice-containing asteroids in which buried ice is occasionally exposed to the heat of the Sun, probably by impacts.  This idea, which seems reasonable but which remains essentially untested, leads us to believe that the orbital distribution of MBCs should be determined jointly by the distribution of ice-containing objects in the main-belt and by the distribution of the asteroid-asteroid collision frequency (related to the local density and other belt parameters).  As for the first quantity, it is reasonable to expect that buried ice is more common in the outer belt than in the inner regions because the rotationally-averaged body-temperature varies with semimajor axis as $a^{-1/2}$.  Evidence for radial compositional gradients has long been recognized in the different distributions of the taxonomic classes, with S (metamorphosed) types more common at smaller $R$ than the C (more primitive) types.  The data and models of thermal stability are, however, consistent with the possibility that \textit{all} outer belt asteroids contain ice.  The ratio of MBCs (on which the ice is temporarily exposed) to outer belt asteroids would then be given roughly by the fraction of the asteroids which experience an excavating collision within the (probably short) lifetime of the exposed ice patch.  Work is underway in Hawaii to begin to determine some of these quantities so that the likely incidence of buried ice can be assessed.

Lastly, note that if water could not be trapped in the hot, young Earth then neither could other, more volatile species such as the noble gases.  Even the outer asteroid belt is not cold enough to trap noble gases in abundance. Sources within more distant, colder cometary reservoirs, probably the Kuiper belt ($T \sim$ 40 K), seem required \cite{owe92}.  The full picture of the delivery of volatiles to the terrestrial planets will probably turn out to be complicated, with multiple sources. \\

\textit{\textbf{Question:} How many MBCs are there? What is their orbital element distribution and what does this tell us about the sources of these bodies?}

\textit{\textbf{Question:} Can we obtain direct evidence (spectroscopy) for the suspected water driver of MBC activity?}

\textit{\textbf{Question:} How are they activated?}

\textit{\textbf{Question:} What fraction of the asteroids as a whole contain ice?}

\textit{\textbf{Question:} What, if anything, can the MBCs tell us about the origin of the Earth's oceans and about terrestrial planet
volatiles in general?}

\section{Comets and their Debris}
\label{transition}
\subsection{Comets Alive, Dormant and Dead}
Objects which are comet-like as judged by their orbits (Tisserand parameters $T_J <$ 3), but which show no evidence for mass loss cannot be classified as comets on physical grounds.   They are sometimes known as Transition Objects (TOs).  \index{Transition objects}The simplest interpretation is that the Transition Objects are comets in which the lack of activity is due to the depletion of near-surface volatiles.  Thermal conduction sets the relevant vertical scale for depletion to the ``skin depth'', of order $\ell \sim (\kappa t)^{1/2}$, where $\kappa$ is the thermal diffusivity of the upper layers and $t$ is the timescale for variation of the Solar insolation.  At least three timescales and three resulting skin depths are relevant (see Table \ref{timescales}, in which I assumed $\kappa$ = 10$^{-7}$ m$^2$ s$^{-1}$ as is appropriate for a powdered dielectric solid).  

\begin{table}
\centering
\caption{Timescales and Skin Depths} \index{Thermal skin depth}
\label{timescales}       
\begin{tabular}{lcc}
\hline\noalign{\smallskip}
Variation & Timescale, $t$ & Skin Depth, $\ell$ [m] \\
\noalign{\smallskip}\hline\noalign{\smallskip}
Diurnal & 10 hr & 0.06 \\
Orbital & 10 yr & 5 \\
Dynamical & 4$\times$10$^5$ yr & 1000 \\
\noalign{\smallskip}\hline
\end{tabular}
\end{table}

The effects of diurnal heating, in particular, can be attenuated by a very modest refractory layer (``mantle'')  just a few centimeters thick.  Direct evidence for this comes from, for example, NASA's Deep Impact mission to comet 9P/Tempel 1, where remote observations have been interpreted as showing a characteristic thickness $\sim$10cm \cite{kad07}.  Mass loss from a comet on which the mantle is much thicker than $\ell$ will be stifled, earning the comet the ``Transition Object'' label. 

Whether or not cometary activity resumes depends upon the long-term stability of the mantle, which itself depends on the dynamical evolution of the comet.  If the mantle lacks cohesion, steady inward drift of the perihelion will lead to increasing temperatures and, eventually, to the ejection of the mantle by gas pressure forces and to the Phoenix-like rebirth of measurable mass loss \cite{ric90}.  [With cohesion, the mantle is potentially much more stable and the mechanism of its failure less easily understood \cite{kuh94}].   Since very thin mantles inhibit sublimation, the mantle formation timescales are probably very short, perhaps comparable to, or even less than, the orbital period \cite{ric90}, \cite{jew02}.  In this simple picture, it is thus likely that the mantles adjust and re-grow as the orbit evolves.

Direct observations of cometary nuclei (comets 1P/Halley, Borrelly, Wild 2 and Tempel 1) confirm the existence of widespread refractory mantles  (e.g. \cite{bas07}) and show that mass loss is channeled through a small number of active areas which, combined, occupy 0.1\% to 10\% of the nucleus surface.  However, other observations throw into doubt the role of mantles in the global control of cometary mass loss.  Most important are measurements of the dust trails of comets.  The dust trail masses, $m_t$, and the cometary mass loss rates, $dm$/$dt$, together define a trail production timescale, $\tau_t$ = $m_t$/($dm$/$dt$).  Separately, dynamical spreading of the trails under the action of planetary perturbations determines the dynamical age of the trail, $\tau_{dyn}$.  Where meaningful measurements of both $\tau_t$ and $\tau_{dyn}$ have been possible, the timescales are found to be very different, with $\tau_t \gg \tau_{dyn}$.  In other words, cometary mass loss at the measured rates cannot supply the trail mass even if continuous over the age of the trail.  This suggests that the trails are not populated by the steady, mantle-choked loss of mass from the nucleus but by some other, more impulsive phenomenon.  Nucleus break-up seems to be the best explanation.  

\begin{figure}
\centering
\includegraphics[height=7cm]{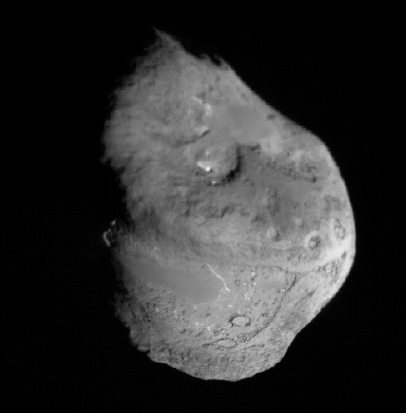}
%
%
\caption{Nucleus of comet Tempel 1.  The visible surface is a refractory mantle (albedo $\sim$4\%) which displays many intriguing landforms, few of which are understood.
Image courtesy of NASA and the Deep Impact team.}
\label{tempel1}       
\end{figure}

Unfortunately, we lack a quantitative understanding of why comets (other than those that are sheared apart by gravity when passing close to planets or the Sun) break up.  Suggested causes include spin-up leading to centripetal disruption \cite{sam86}, high internal gas pressures caused by sublimating supervolatiles (Samarasinha's most enjoyable ``bomb'' model \cite{sam01}), impact with unseen interplanetary debris and disruption by thermally induced stresses.  All of these ideas verge on the fantastic, with the exception of centripetal disruption, which is a natural outcome of torques applied to the nucleus by non-uniform outgassing.  I know of no data to suggest a relationship between nucleus spin rate and break-up but this could be simply because there are too few relevant nucleus spin measurements (i.e. ``absence of evidence'' should not be construed as ``evidence of absence'', as far as the spin vs. break-up connection is concerned).  The lack of understanding is disconcerting given the potential importance of break-up in determining the fates of small bodies.  \\

\textit{\textbf{Question:} How many TOs are there?  The number of TOs relative to the number of active comets will tell us the
ratio of the outgassing to the dynamical lifetimes of these bodies.}

\textit{\textbf{Question:} What is their orbital element distribution and what does this tell us about the sources of these bodies?}

\textit{\textbf{Question:} Do  \textit{all} comets evolve into TOs or do some proceed directly to disintegrate into debris streams?}

\textit{\textbf{Question:} Are the TOs dead or dormant, or both?  In other words, is the ice depleted down to the core, or just down
to a few times the thermal skin depth?}

\textit{\textbf{Question:} How do TOs die?  Are their lifetimes limited by impact with the planets or the Sun, by dynamical ejection,
or by a physical process such as break-up?}

\subsection{Damocloids}
\label{damocloids}
The Damocloids\index{Damocloids} are a subset of the Transition Object \index{Transition objects}class, named after the prototype object (5335) Damocles.  They are defined by having a point-source appearance and 
$T_J <$ 2 \cite{jew05}.  At the time of writing (2007 May 22), 36 objects meet this definition.  The orbits of the Damocloids are statistically similar to the orbits of Halley family and long-period comets (e.g. many Damocloid orbits are retrograde), rather than with the Jupiter family.  The association is further strengthened by the fact that some bodies originally classified as Damocloids have, since discovery, been found to show weak comae.  Damocloids, then, are the inactive nuclei of comets recently emplaced in the planetary region of the Solar system from a source probably located in the inner Oort Cloud \cite{don04}.  Curiously, although their dynamical and evolutionary histories have been quite different from those of the short-period comets, the surface properties of the two classes of comet nucleus are indistinguishable
 \cite{jew05}. \\

\textit{\textbf{Question:}  How many Damocloids exist and what is the ratio of Damocloids to Halley Family Comets?}

\textit{\textbf{Question:} What is the size distribution of the Damocloids?}

\textit{\textbf{Question:} Is there evidence that some Damocloids might be intrinsically refractory bodies (asteroids) ejected into 
the Oort Cloud and then scattered back to the inner Solar system, as has been suggested for 1996 PW \cite{wei97}.}

\textit{\textbf{Question:} Do any Damocloids carry Ultra-Red matter?  The published sample does not, but the published sample is small.}

\begin{figure}
\centering
\includegraphics[height=7cm]{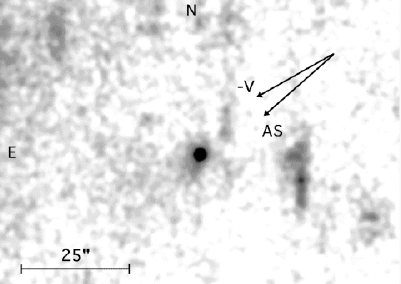}
%
%
\caption{``Asteroid'' 2005 WY25 at 1.6 AU showing ultra-weak outgassing in a 1500 second, R-band image from the UH 2.2-m telescope on UT 2004 March 20.  The mass loss rate inferred from the coma is very uncertain, but of the order 10 g s$^{-1}$.  2005 WY25 is the likely parent
of the Phoenicid meteor stream, and a probable fragment of comet D/1819 W1 (Blanpain).
Figure from \cite{jew06}.}
\label{2005wy25}       
\end{figure}

\subsection{Meteor Stream Parents} \index{Meteor stream parent}
One fate for cometary nuclei is to disintegrate, forming a trail of solid debris particles that can be detected remotely from their
thermal emission (\cite{rea07}) and optical signatures (\cite{ish02}) or directly, if their orbits intersect that of the Earth and produce a meteor stream.  Recent work has given a boost to the study of meteor streams and the parent bodies which produce them.  Significantly, some of the parent objects have now been identified with confidence.  One surprise is that not all the parents are comets: some streams seem to result from the breakup of bodies, like (3200) Phaethon, which are dynamically asteroids.  

\begin{figure}
\centering
\includegraphics[height=8cm]{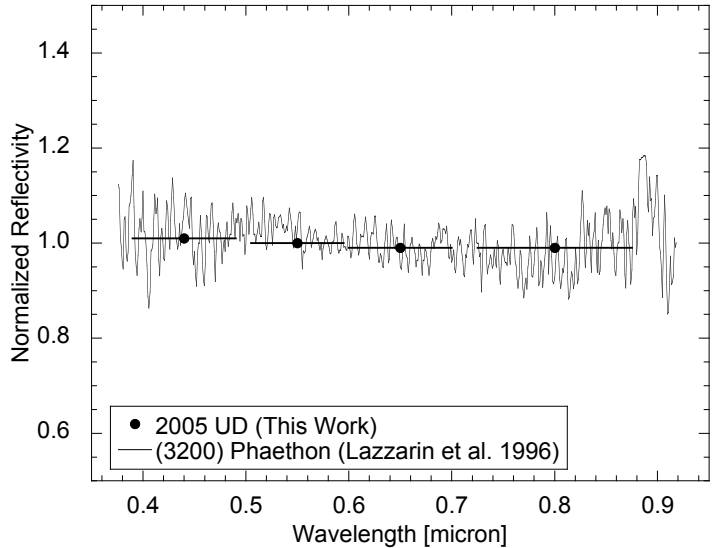}
%
%
\caption{Reflection spectrum of Asteroid 2005 UD (points) compared with the spectrum of dynamically related object (3200) Phaethon.  Both
objects are unusual in showing spectra slightly bluer than the Sun in reflected light.
Figure from \cite{jh06}.}
\label{2005ud}       
\end{figure}

Asteroid (3200) Phaethon ($T_J$ = 4.508) has orbital elements similar to those of the Geminid meteors. On this
basis, Phaethon was long-ago proposed as a likely Geminid stream parent \cite{whi83}.  Recently discovered asteroid 2005 UD ($T_J$ = 4.504)  has very similar orbital elements and is probably related both to Phaethon
and to the Geminids \cite{oht06}.  The albedo of Phaethon has been measured as 0.11$\pm$0.02 and the diameter as 4.7$\pm$0.5 km \cite{gre85}. If the albedo of 2005 UD is the same, then its diameter must be only 1.3$\pm$0.1 km \cite{jh06}.   Sensitive, high resolution imaging observations provide no evidence for on-going mass loss, either from Phaethon \cite{hsi05} or from 2005 UD \cite{jh06}, above 
the level of $\sim$ 10$^{-2}$ kg s$^{-1}$.   The age of the Geminid stream estimated from
dynamical considerations is about 1000 yr \cite{wil93}.  In 1000 yr, mass loss at 10$^{-2}$ kg s$^{-1}$ would give a stream mass $M_s \sim$ 3$\times$10$^8$ kg, whereas the mass has been independently estimated at $M_s \sim$ 1.6$\times$10$^{13}$ kg \cite{hug89}.  This huge discrepancy (a factor $\sim$10$^5$) indicates that meteor stream formation must be episodic or even catastrophic, not steady-state.  

Both Phaethon and 2005 UD show slightly blue optical reflection spectra of unknown origin (Figure \ref{2005ud}). Blue reflection spectra are uncommon (only 1 out of $\sim$23) amongst the near-Earth objects  \cite{jh06}, \cite{kin07}.
There is speculation that the blue color could reflect thermally altered minerals on these bodies at the 
high temperatures (perhaps 740 K) resulting from their small perihelion distances ($q \sim$ 0.14 AU) \cite{lic07}.  Likewise, the high mean density of the Geminids 
($\rho \sim$ 2900 kg m$^{-3}$) is also unusual and has been suggested to result from compaction associated with loss of volatiles \cite{kas06}. On the other hand, observational support for thermal desorption is lacking: careful spectroscopic measurements of the $Na$/$Mg$ ratio show that the Geminids are not compositionally different from other meteoroids with much larger
$q$ $\sim$1 AU \cite{kas06}.  A reasonable guess is that the Geminid meteors, Phaethon and 2005 UD (and probably other macroscopic bodies yet to be found) are products of the recent breakup of a precursor body  \index{Meteor stream parent}but the nature of the precursor and the cause of the breakup have yet to be determined.\\

\textit{\textbf{Question:} Are all objects with small $q$ necessarily blue as a result of thermal alteration?}

\textit{\textbf{Question:} What kind of body was the Geminid precursor?}

\textit{\textbf{Question:}  What caused the precursor to breakup? Thermal stresses?  Internal gas pressure forces?  Spin-up by outgassing or radiation forces?}

\textit{\textbf{Question:}  How does the rate at which mass is input to the interplanetary medium by catastrophic disruption of meteor stream parents compare with the rates from cometary sublimation and from asteroid-asteroid collisions in the main-belt?}  \\

\section{Epilogue}
A reasonable conclusion to be drawn from this chapter is that planetary astronomy is a most active and revitalized field.  Key advances are being made in the determination of the contents of the Solar system, with the discovery of new populations of bodies and the unveiling of links between populations that were, until recently, unsuspected. These new observational results, combined with the rising power of computers, together motivate exciting new conjectures for the origin and evolution of the Solar system.  \\

\textbf{Acknowledgements}

I thank Toshi Kasuga, Pedro Lacerda, Ingrid Mann, Nuno Peixinho, Rachel Stevenson and Bin Yang for comments on this manuscript and NASA's Planetary Astronomy and Origins programs for support of the work described herein.




\printindex
\end{document}